\pdfoutput=1
\documentclass{article}
    \usepackage[preprint]{neurips2025}

\usepackage{times}
\usepackage{latexsym}

\usepackage[T1]{fontenc}
\usepackage[utf8]{inputenc}

\usepackage{microtype}

\usepackage{inconsolata}

\usepackage{graphicx}
\usepackage{subcaption}
\usepackage{booktabs} % for professional tables
\usepackage{mdframed}
\usepackage{colortbl}
\usepackage{multirow}
\usepackage{float}
\usepackage{pifont}
\usepackage{diagbox}
\usepackage{url}

\usepackage{hyperref}
\usepackage{amsmath}
\usepackage{amssymb}
\usepackage{mathtools}
\usepackage{amsthm}
\usepackage{array}
\usepackage{multirow}
\usepackage[capitalize,noabbrev]{cleveref}
\definecolor{lightergray}{gray}{0.90}
\newcolumntype{a}{>{\columncolor{lightergray}}c}
\newcolumntype{x}{>{\columncolor{lightergray}}l}
\newcommand{\modelname}{\textsc{VoiceStar}}
\newcommand{\pmrope}{\textsc{PM-RoPE}}
\newcolumntype{L}[1]{>{\arraybackslash}p{#1}}

\usepackage{wrapfig}
\usepackage{xcolor}
\usepackage{makecell}
  
\newcommand{\pyp}[1]{{\color{purple}[Puyuan: #1]}}

\newcommand{\daniel}[1]{{\color{teal}[Daniel: #1]}}
\definecolor{myLightGray}{gray}{0.40} % or use gray!50, adjust as needed
\renewcommand{\daniel}[1]{}
\renewcommand{\pyp}[1]{}
\let\citet\cite
\let\citep\cite
\title{\modelname: Robust Zero-Shot Autoregressive TTS with Duration Control and Extrapolation}

\author{Puyuan Peng$^1$ \qquad Shang-Wen Li \qquad Abdelrahman Mohamed$^2$ \qquad \textbf{David Harwath}$^1$ \\
        $^1$The University of Texas at Austin \qquad Rembrand$^2$\\
\texttt{pyp@utexas.edu}}

\begin{document}
\maketitle
\begin{abstract}
We present \modelname, the first zero-shot TTS model that achieves both output duration control and extrapolation. \modelname~is an autoregressive encoder-decoder neural codec language model, that leverages a novel Progress-Monitoring Rotary Position Embedding (\pmrope) and is trained with Continuation-Prompt Mixed (CPM) training. \pmrope~enables the model to better align text and speech tokens, indicates the target duration for the generated speech, and also allows the model to generate speech waveforms much longer in duration than those seen during. CPM training also helps to mitigate the training/inference mismatch, and significantly improves the quality of the generated speech in terms of speaker similarity and intelligibility. \modelname~outperforms or is on par with current state-of-the-art models on short-form benchmarks such as Librispeech and Seed-TTS, and significantly outperforms these models on long-form/extrapolation benchmarks (20-50s) in terms of intelligibility and naturalness. Code and model weights will be open-sourced. Code and model: \href{https://github.com/jasonppy/VoiceStar}{github.com/jasonppy/VoiceStar}. Audio samples: \href{https://jasonppy.github.io/VoiceStar_web/}{jasonppy.github.io/VoiceStar\_web/}
\end{abstract}

\section{Introduction}
Neural codec language models (NCLMs) have rapidly become a state-of-the-art method for text-to-speech (TTS) generation. 
These models use neural network audio codecs~\cite{Zeghidour2021SoundStreamAE,Defossez2022HighFN} to tokenize speech waveforms into sequences of discrete symbols representing temporal frames. 
\begin{wrapfigure}{r}{0.4\textwidth}  % "r" for right, width can be adjusted
\vspace{-0.2in}
    \centering
    \includegraphics[width=\linewidth]{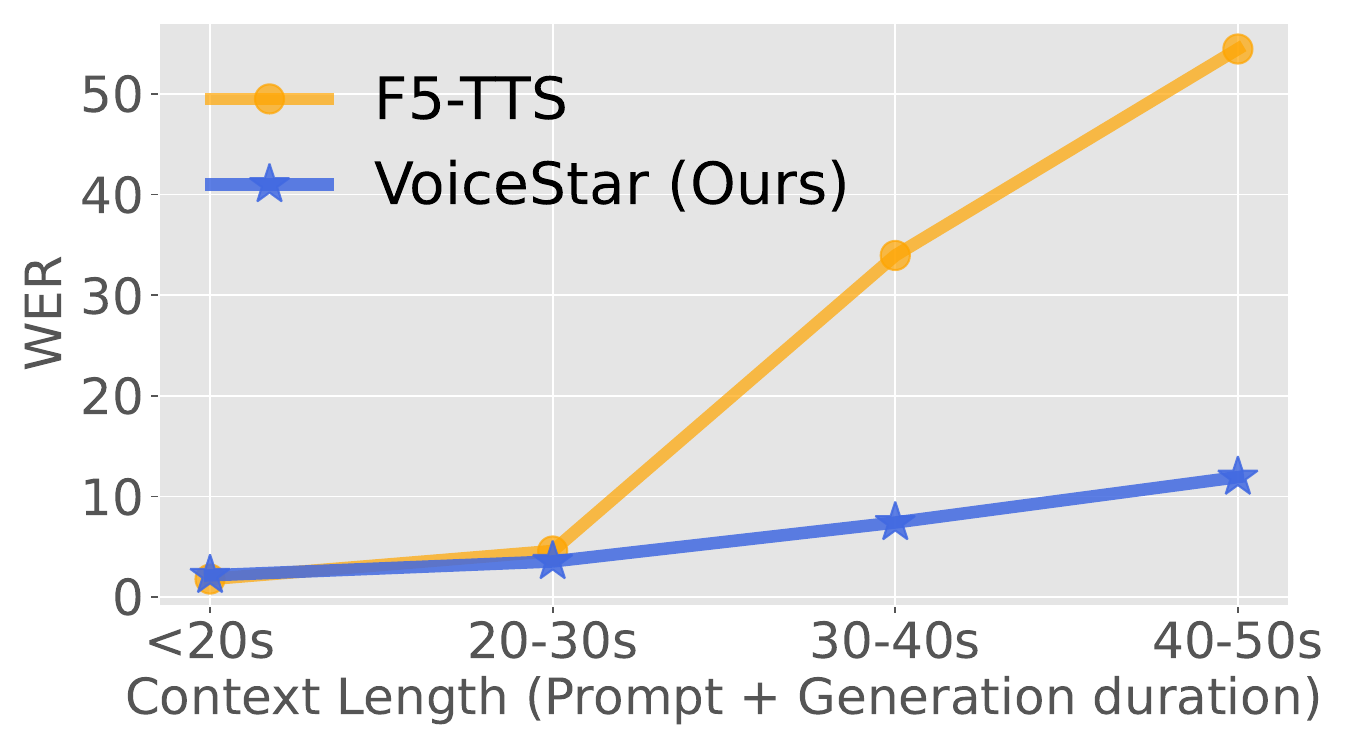}
    \caption{WER comparison between our \modelname~and F5-TTS~\cite{f5tts} under different context lengths. Both models are trained with maximal context length of 30 seconds.}
    \label{fig:teaser}
    \vspace{-0.25in}
\end{wrapfigure}
Next, a Transformer~\cite{Vaswani2017AttentionIA} language model is used to autoregressively model these token sequences. The success of this approach is due to combination of the modeling power of Transformer language models and the ease of reconstructing high-fidelity waveforms from the generated token sequences.
However, current NCLM-based TTS models fall short in several important ways, specifically their lack of fine-grained controllability (especially for duration control) and their inability to extrapolate to sequence lengths much longer than those seen during training. 

In this paper, we propose solutions for these problems, and also propose  several other novel techniques for improving the quality of speech generated by NCLM TTS models.

Specifically, we note that current NCLM-based TTS models~\cite{Wang2023NeuralCL,chen2024valle2neuralcodec,voicecraft} do not explicitly model the alignment between their input text and speech sequences, and instead simply concatenate these sequences and put the onus on the model to learn the alignment via standard positional encodings. Furthermore, standard positional encodings do not easily enable a user to specify the desired sequence length of a generation at inference time. 

To fix these flaws, we propose to use a Transformer encoder-decoder architecture coupled a novel Progress-Monitoring Rotary Position Embedding (\pmrope). This provides the model with a form of flat-start alignment between text and speech tokens from the very beginning of training. The \pmrope embeddings also encode the desired sequence length for the generated speech, which implicitly informs the model at each timestep how far along the generation has progressed relative to this target length. Furthermore, we also note that existing NCLM-based TTS models treat voice cloning TTS as speech continuation, which at test time can entangle the reference speaker's voice characteristics with the prosody present in the reference utterance. At inference time, we may only wish to clone the speaker's vocal characteristics and let the model infer what the prosody or emotional delivery should be conditioned on the text that should be synthesized. To encourage this disentanglement, during training we perform random prompt mixing: sometimes the model is trained to perform continuation of an utterance, but other times we sample a random (different) utterance from the same speaker to serve as the reference prompt. We call this method continuation-prompt mixed training (CPM training). The use of CPM training additionally allows us to apply data augmentation techniques, such as speed perturbation, which we show improves intelligibility. %; 2) prompt repetition during inference improves speaker similarity.

% \begin{figure}
%     \centering
%     \includegraphics[width=1\linewidth]{src/figs/context_length_teaser.pdf}
%     \caption{WER comparison between our \modelname~and F5-TTS on generation under different context lengths (context length is defined as prompt + generation duration). Both models are trained with maximal context length of 30 seconds.}\label{fig:teaser}
% \end{figure}

To summarize, our contributions are as follows:
\begin{enumerate}
    \item We propose Progress-Monitoring Rotary Position Embeddings, or \pmrope, that leads to robust and duration controllable NCLM-based TTS, which further unlocks the capability to generate utterances much longer than those seen during training;
    \item We propose continuation-prompt mixed training, or CPM training, which improves the intelligibility and naturalness of NCLM-based TTS models;
    \item We propose two additional techniques that positively impact the performance of NCLM-based TTS models: 1) prompt speed perturbation during training; 2) prompt repetition during inference.
\end{enumerate}
Combining the proposed techniques, \textbf{our model, \modelname, is the first zero-shot TTS model with duration control and length extrapolation capabilities}. VoiceStar achieves performance better than or on par with other state-of-the-art models on existing short-form benchmarks including Seed-TTS-eval and Librispeech, and significantly outperforms current SotA models on long-form/extrapolation benchmarks.

\section{Related Work}
\textbf{Neural Codec Language Models}
Pioneered by~\cite{Lakhotia2021OnGS,Borsos2022AudioLMAL,Kreuk2022AudioGenTG,Wang2023NeuralCL,Kharitonov2023SpeakRA,Borsos2023SoundStormEP}, NCLMs have become one of two state-of-the-art approaches for audio generation, the other being flow-matching/diffusion models. 
\begin{wraptable}{r}{0.65\textwidth} 
\vspace{-0.1in}
\centering
\caption{Conceptual comparison of \modelname~with other models. 
  AR stands for autoregressive and NAR stands for non-autoregressive. Extrapola. stands for extrapolation.
}
\resizebox{1\linewidth}{!}{
\begin{tabular}{lccccc}
    \toprule
    Model & Paradigm & \makecell[c]{Open\\Source} & \makecell[c]{Voice\\Cloning} & \makecell[c]{Duration\\Control} & \makecell[c]{Extrapola.} \\
    \midrule
    VALL-E~\cite{Wang2023NeuralCL} & AR+NAR &  & \checkmark &  &  \\
    Voicebox~\cite{Le2023VoiceboxTM} & NAR &  & \checkmark & \checkmark &  \\
    MaskGCT~\cite{Wang2024MaskGCTZT} & NAR & \checkmark & \checkmark & \checkmark &  \\
    F5-TTS~\cite{f5tts} & NAR & \checkmark & \checkmark & \checkmark &  \\
    VAT~\cite{Battenberg2024VeryAT} & AR &  &  &  & \checkmark \\
    VoiceCraft~\cite{voicecraft} & AR & \checkmark & \checkmark &  &  \\
    CosyVoice~\cite{Du2024CosyVoiceAS,Du2024CosyVoice2S} & AR & \checkmark & \checkmark &  &  \\
    Llasa~\cite{Ye2025LlasaST} & AR & \checkmark & \checkmark &  &  \\
    \modelname~(Ours) & AR & \checkmark & \checkmark & \checkmark & \checkmark \\
    \bottomrule
\end{tabular}
}
\label{tab:comparison}
\vspace{-0.2in}
\end{wraptable}
Due to the strong in-context learning ability of Transformer LMs, NCLMs are a particularly effective approach for zero-shot voice-cloning TTS~\cite{Wang2023NeuralCL}. In this setting, a model must clone the vocal characteristics of a reference speaker that was unseen during training, using several seconds of reference speech provided as a prompt at inference time.

\textbf{Enhancing the Robustness of NCLM.}
Despite their typically strong performance, NCLMs are known to have robustness issues, such as skipping words, inserting extra words, repeating words, and inserting unnaturally long silences~\cite{Wang2023NeuralCL,voicecraft}. Several papers have proposed to address these issues from the angle of text-speech alignment. Specifically,~\cite{Song2024ELLAVSN,Han2024VALLERR} use an external forced alignment model to tightly couple the text prompt and generated speech, while~\cite{Du2024VALLTDG} makes use of a transducer architecture to implicitly learn text-speech alignment.
~\cite{Wang2024AttentionConstrainedIF} uses a constrained attention mechanism to enforce monotonic text-speech alignment.~T5-TTS~\cite{Neekhara2024ImprovingRO} loads weights from a textual encoder-decoder T5 model and introduces an auxiliary loss to encourage monotonic alignment in the text-speech cross-attention weights. Also, unlike speech-continuation training, T5-TTS uses a separate utterance drawn from the same speaker as the prompt during training. This differs from our proposed CPM training in two ways: 1) in addition to the reference speech, we also append the transcript to the text that is to be generated; 2) we stochastically mix these same-speaker-different-utterance prompts with utterance continuation-based prompting, which we show in our ablation experiments to be better than either style of prompting on its own.
VAT~\cite{Battenberg2024VeryAT} also uses an encoder-decoder architecture, and proposes a T5-like relative position embedding to enhance the text-speech alignment. Notably, VAT also achieves extrapolation similarly to our model, however it cannot simultaneously control the duration of the output as our model can. Another crucial difference between VoiceStar and VAT is that VAT is a conventional multi-talker TTS system that can only generate speech in the voices of speakers seen during training, and is not capable of zero-shot voice-cloning TTS.

Note that in addition to improving the text-speech alignment in NCLMs, there are other works that have tried other methods to improve robustness, such as using multi-scale generation to address recency bias~\cite{Guo2024SpeakingFC}, enforcing disentanglement of speech attributes~\cite{jiang2024megatts}, incorporating classifier-free guidance~\cite{Wang2024SSRSpeechTS}, using chain-of-thought prompting~\cite{Xin2024RALLERC}, and reinforcement learning from human/AI feedback~\cite{Chen2024EnhancingZT,Hu2024RobustZT,Hussain2025KoelTTSEL}.

\textbf{Adding new capabilities to NCLMs.}
\cite{Zhang2023SpeakFL,Zhu2024GenerativePS} extended NCLM-based TTS to the multilingual case.~\cite{Wang2023VioLAUC,Wang2023LauraGPTLA,Maiti2023VoxtLMUD,Wu2024SpeechComposerUM,Wang2025MetisAF} propose multi-task learning for generation and recognition tasks. \cite{Guo2022PromptttsCT,Yang2023InstructTTSME,Liu2023PromptStyleCS,Ji2023TextrolSpeechAT,Leng2023PromptTTS2D,Lyth2024NaturalLG,anuj2025} adapt NCLMs to style-controlled speech synthesis.
\cite{Defossez2024MoshiAS,Ma2024LanguageMC,Fang2024LLaMAOmniSS,Yu2024SALMONNomniAC,Xu2024EnablingRC,Zhang2024IntrinsicVoiceEL,Zhang2025LLMEnhancedDM} propose NCLM-based models for real-time interactive voice assistants. \cite{Lajszczak2024BASETL,Wang2024MaskGCTZT,Ye2025LlasaST} investigate scaling up NCLM-based models.
~\cite{Nishimura2024HALLEHN} also focuses on long-form generation, but our work differs from theirs in that~\cite{Nishimura2024HALLEHN} specifically trains on long-form speech, while our model is trained on short-form speech only, yet we show that it can generalize to long-form speech.

\textbf{Diffusion/flow-based TTS and Scaling TTS models.} Diffusion/flow-based models represent another popular approach to zero-shot TTS \daniel{a sentence about the difference between NCLM and Diffusion based approach?}. With masked reconstruction training, \cite{Shen2023NaturalSpeech2L,Le2023VoiceboxTM,Kim2023PFlowAF} show that duffision/flow-based models can achieve SotA performance in zero-shot TTS. \cite{Lee2024DiTToTTSDT} employs the DiT~\cite{Peebles2022ScalableDM} architecture with a global duration predictor to improve the performance of diffusion-based TTS.
\cite{Li2024DMOSpeechDM} distills efficient diffusion models via direct metric optimization. E2-TTS~\cite{Eskimez2024E2TE} simplifies the model pipeline by removing phoneme duration prediction and grapheme-to-phoneme modules. F5-TTS~\cite{f5tts} scales E2-TTS to 100k hours of bilingual (English and Mandarin) speech. \cite{Vyas2023AudioboxUA} also scales a flow-based model to allow for a variety of capabilities such as style control and general audio generation. Finally,~\cite{Du2024CosyVoiceAS,Guo2024FireRedTTSAF,Du2024CosyVoice2S} proposed hybrid models that combine NCLM and flow-matching.

\section{Method} 
\begin{figure*}
    \centering
    \includegraphics[width=1\textwidth]{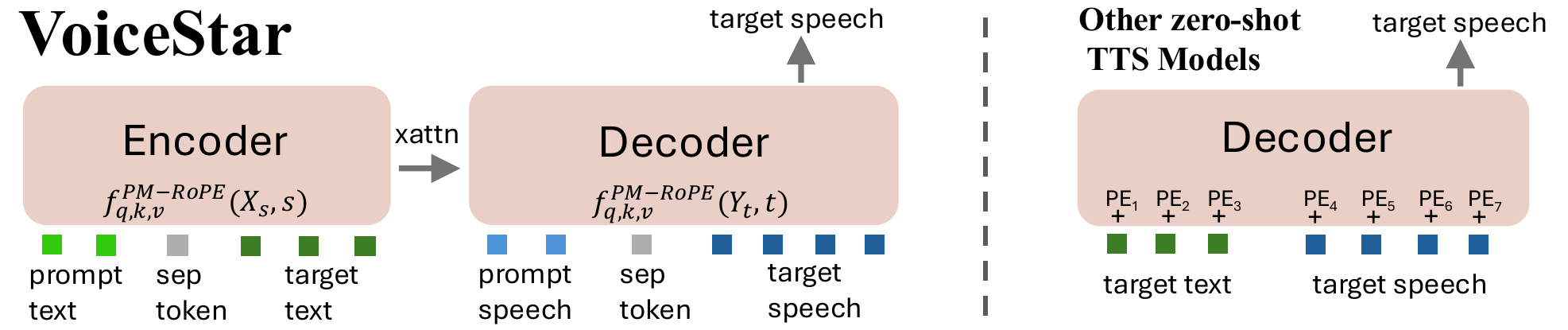}
    \caption{Left: The architecture of \modelname. Right: the common general architecture for zero-shot TTS models, such as VALL-E (AR part), VoiceCraft, CosyVoice, FireRedTTS, Llasa etc. \modelname~differs from them in three aspects: 1) it uses an encoder-decoder architecture with PM-RoPE to provide text-speech alignment, duration control, and extrapolation capability; 2) it uses prompt-continuation mixed training to mitigate the training/inference mismatch and enhance robustness. Our speech tokenizer based on Encodec~\cite{Defossez2022HighFN} uses multiple codebooks and we apply the delay pattern~\cite{musicgen} to them, which is not depicted in this figure for simplicity.}
    \label{fig:model}
\end{figure*}

The architecture of \modelname~along with a typical decoder-only, zero-shot TTS architecture are shown in Fig.~\ref{fig:model}.
\modelname~adopts an encoder-decoder architecture, where the encoder takes as input IPA phonemes derived from the text transcript, and the decoder autoregressively predicts speech tokens produced by an Encodec model\cite{Defossez2022HighFN,voicecraft}. Text and speech inputs are separated with special tokens to indicate reference and target, and are positioned using PM-ROPE, enabling controllable, zero-shot TTS with duration extrapolation capability. The following two sections describe the key differences between our model and prior works in detail. 
\subsection{Progress-Monitoring RoPE}
We first briefly describe the vanilla dot-product attention mechanism~\cite{bahdanau2016neuralmachinetranslationjointly}, then describe the Rotary Position Embedding (RoPE)~\cite{su2023roformerenhancedtransformerrotary}. Next, we describe our proposed extension of RoPE to Progress-Monitoring RoPE (\pmrope) which provides better text-speech alignment, informs the model of the target output duration, and enables inference-time duration extrapolation.

\textbf{Dot-product attention.} We use the commonly used ``source-target'' terminology throughout to describe different attention mechanisms. Note that target and source can reference to the same sequence, in which case attention refers to self-attention.
Denote the source token embedding at position $s$ as $X_s \in \mathbb{R}^D, 1 \leq s \leq S$ and the embedding of target token at position $t$ as $Y_t \in \mathbb{R}^D, 1 \leq t \leq T$. The embeddings first get projected into key, query, and value vectors as
% \begin{align}
%     &K_s = f_k(X_s, s) := W_k X_s, \label{eq:k_projection} \\
%     &Q_t = f_q(Y_t, t) := W_q Y_t, \\
%     &V_s = f_v(X_s, s) := W_v X_s \label{eq:v_projection}
% \end{align}
\begin{equation*}
    K_s = f_k(X_s, s) := W_k X_s, \quad Q_t = f_q(Y_t, t) := W_q Y_t, \quad V_s = f_v(X_s, s) := W_v X_s
\end{equation*}

Where $W_k$, $W_q$, and $W_v$ are learnable matrices. 

We denote the inner product operator as $\langle \cdot,\cdot \rangle$. Also let the attention weight from target token at position $t$ to source token at position $s$ be $W_{t,s}$, and the output of the attention layer at position $t$ be $O_t$, which are calculated according to: $W_{t,s} = \exp\left (\frac{\langle f_q(Y_t, t),  f_k(X_s, s)\rangle}{\sqrt{D}}\right )/\sum_{s'}\exp\left (\frac{\langle f_q(Y_t, t),  f_k(X_{s'}, s')\rangle}{\sqrt{D}}\right )$ and $O_t = \sum_{s'} W_{t,s'} V_{s'}$.
% \begin{align}
%     &W_{t,s} = \frac{\exp\left (\frac{\langle f_q(Y_t, t),  f_k(X_s, s)\rangle}{\sqrt{D}}\right )}{\sum_{s'}\exp\left (\frac{\langle f_q(Y_t, t),  f_k(X_{s'}, s')\rangle}{\sqrt{D}}\right )} \label{eq:weight}\\
%     &O_t = \sum_{s'} W_{t,s'} V_{s'}\label{eq:output}
% \end{align}
% \begin{equation}
%     W_{t,s} = \frac{\exp\left (\frac{\langle f_q(Y_t, t),  f_k(X_s, s)\rangle}{\sqrt{D}}\right )}{\sum_{s'}\exp\left (\frac{\langle f_q(Y_t, t),  f_k(X_{s'}, s')\rangle}{\sqrt{D}}\right )}, \quad O_t = \sum_{s'} W_{t,s'} V_{s'}\label{eq:output}
% \end{equation}

The attention machenism is a powerful way to provide contextualization, but it does not explicitly include token position information. Therefore, various position embedding methods have been proposed to address this issue~\cite{vaswani2023attentionneed,raffel2023exploringlimitstransferlearning,su2023roformerenhancedtransformerrotary,press2022trainshorttestlong}.

\textbf{RoPE}~\cite{su2023roformerenhancedtransformerrotary}. Rotary Position Embedding (RoPE) redefines the projection functions $f_q(Y_t, t)$ and  $f_k(X_s, s)$ such that their inner product is a function $g()$ of the corresponding target and source token embeddings, and their relative position, which can be formally written as:
\begin{equation*}
    \langle f^{\textsc{RoPE}}_q(Y_t, t),  f^{\textsc{RoPE}}_k(X_s, s)\rangle = g(Y_t,  X_s, t-s)
\end{equation*}

For simplicity, we first explain RoPE in the 2-dimensional case. The 2-D rotation matrix $R(\gamma)$ is defined as:
\begin{equation*}
    R(\gamma) = \left(
    \begin{array}{cc}
        \cos{\gamma}& -\sin{\gamma}  \\
        \sin{\gamma}&\cos{\gamma} 
    \end{array}
    \right)
\end{equation*}
RoPE applies this rotation matrix to key and query calculation according to:
% \begin{align}
%     &f^{\textsc{RoPE}}_k(X_s, s) = R(s\theta) W_k X_s \\
%     &f^{\textsc{RoPE}}_q(Y_t, t) = R(t\theta) W_q Y_t
% \end{align}
\begin{equation*}
    f^{\textsc{RoPE}}_k(X_s, s) = R(s\theta) W_k X_s, \quad f^{\textsc{RoPE}}_q(Y_t, t) = R(t\theta) W_q Y_t
\end{equation*}
Where $\theta$ is a hyperparameter that specifies the per position rotation angle. The inner product between the key and the query becomes:
% \begin{align*}
%     &\langle f^{\textsc{RoPE}}_q(Y_t, t),  f^{\textsc{RoPE}}_k(X_s, s)\rangle \\
%     &= Y_t^\top W_q^\top R((t-s)\theta) W_k X_s
% \end{align*}
\begin{equation*}
    \langle f^{\textsc{RoPE}}_q(Y_t, t),  f^{\textsc{RoPE}}_k(X_s, s)\rangle= Y_t^\top W_q^\top R((t-s)\theta) W_k X_s
\end{equation*}
Observe that due to the application of the rotation matrix, the inner product is a function of $Y_t$, $X_s$, and their relative position within the input sequence $t-s$.

The value vector is calculated the same as in vanilla attention, i.e. $f^{\textsc{RoPE}}_v(X_s, s) := f_v(X_s, s)$. Also the attention weights $W_{t,s}$ and overall attention output $O_t$ are calculated similarly to vanilla attention. An important property of \textsc{RoPE} is recency bias, i.e. the inner product between two tokens decreases as the distance between them increases, and therefore for any given position, nearby tokens will have a larger contribution to the attention output for that position. 
For higher dimensional spaces, RoPE divides the space into D/2 2-dimensional spaces, and applies 2-dimensional rotation matrices with different $\theta$s to each 2-dimensional space; more details can be found in Appendix~\ref{sec:app-detail}.

\textbf{Progress-Monitoring RoPE (\pmrope)}. While RoPE has been widely used in textual decoder-only LLMs, speech-text LMs utilize two distinct input sequences which often are implicitly aligned. We therefore propose a novel extension to RoPE, namely Progress-Monitoring RoPE (\pmrope). When combined with an encoder-decoder NCLM, \pmrope~brings three benefits: 1) improved text-speech alignment, 2) duration control during generation, and 3) test-time extrapolation. Mathematically, \pmrope~introduces a simple change to $f_k$ and $f_q$:
% \begin{align}
%     &f^{\textsc{PM-RoPE}}_k(X_s, s) = R\left (\frac{s}{S}N\theta\right ) W_k X_s \\
%     &f^{\textsc{PM-RoPE}}_q(Y_t, t) = R\left (\frac{t}{T}N\theta\right ) W_q Y_t
% \end{align}
\begin{equation*}
    f^{\textsc{PM-RoPE}}_k(X_s, s) = R\left (\frac{s}{S}N\theta\right ) W_k X_s, \quad f^{\textsc{PM-RoPE}}_q(Y_t, t) = R\left (\frac{t}{T}N\theta\right ) W_q Y_t
\end{equation*}
% $$f^{\textsc{PM-RoPE}}_q(Y_t, t) = R\left(\frac{\boldsymbol{t}}{\boldsymbol{T}} \boldsymbol{N} \theta\right) W_q Y_t$$
where instead of using \textit{positions} $s$ and $t$, we measure the \textit{fractional progress} of $s$ towards some maximum value $S$, and the progress of $t$ towards some maximum value $T$, i.e. $\frac{s}{S}$ and $\frac{t}{T}$. Specifically, $S$ is the total sequence length of the source tokens, and $T$ is the (desired) total sequence length of the target tokens. $N$ is a hyperparameter that rescales the rotation angle, which can also be interpreted as the \textit{pseudo} total sequence length. The inner product between a key vector and a value vector becomes:
% \begin{align}
%     &\langle f^{\textsc{PM-RoPE}}_q(Y_t, t),  f^{\textsc{PM-RoPE}}_k(X_s, s)\rangle \\
%     &= Y_t^\top W_q^\top R\left (\left (\frac{t}{T}-\frac{s}{S}\right )N\theta\right ) W_k X_s
% \end{align}
\begin{equation*}
    \langle f^{\textsc{PM-RoPE}}_q(Y_t, t),  f^{\textsc{PM-RoPE}}_k(X_s, s)\rangle = Y_t^\top W_q^\top R\left (\left (\frac{t}{T}-\frac{s}{S}\right )N\theta\right ) W_k X_s
\end{equation*}
Note that the inner product is a function of $Y_t$, $X_s$ and their \textit{relative progress} $\frac{t}{T}-\frac{s}{S}$. Similar to \textsc{RoPE}, \pmrope~also enjoys the long-term decay property, i.e. the inner product between two tokens decreases as the difference between their progresses increases.

In our encoder-decoder NCLM (the left hand side in Fig.~\ref{fig:model}), the encoder takes as input a phonemized text transcript, and the decoder predicts acoustic tokens in an autoregressive fashion. We treat the phonetic tokens as the source tokens $X$, and the acoustic tokens as the target tokens $Y$.
When \pmrope~is applied to the decoder's self-attention mechanism, it provides information about the location of the current acoustic token within the overall target sequence length, which enables output duration control - during autoregressive generation, when the progress of the target token sequence reaches 100\% (i.e. $\frac{t}{T} = 1$, where the value of $T$ is specified in advance by the user), the model should learn to emit an end-of-generation token. 
When \pmrope~is applied to the cross-attention connecting the encoder to the decoder (acoustic tokens attending to phonetic tokens), it provides information about the temporal alignment between the two modalities.  This is helpful during training because it provides a flat-start initial alignment to the untrained model, since an acoustic token's cross-attention will be concentrated at the phonetic tokens occupying the same relative position within the phonetic sequence, i.e. where the difference $\frac{t}{T} - \frac{s}{S}$ is small. See Appendix~\ref{sec:alignment} for more details. %Further, the enhances the duration control by informing decoder the progress in the phoneme tokens. 
\begin{wrapfigure}{r}{0.5\textwidth}  % "r" for right, width can be adjusted
    \vspace{-0.1in}
    \centering
    \includegraphics[width=\linewidth]{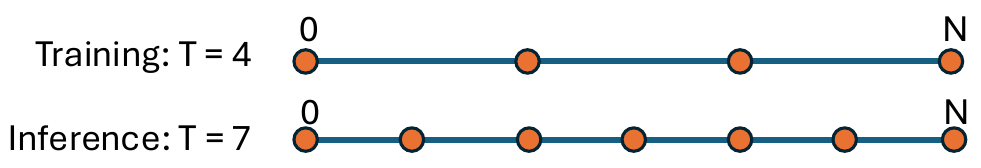}
    \caption{An example on how \pmrope~turns extrapolation into interpolation: during training, the maximal training sequence length is 4, and during inference the target length is 7. The positional encodings for both can be expressed as sampling points inside the same interval $[0, N]$.}
    \label{fig:extrapolation}
\end{wrapfigure}
We found it to be necessary to apply \pmrope~to the encoder's phonetic input sequence to achieve extrapolation, which we believe is due to the fact that it allows both the encoder and decoder sequences to have a fixed pseudo sequence length $N$. Therefore, we can extrapolate to longer phonetic or acoustic token sequences by sampling a larger number of points in the interval $[0,N]$ and interpolating between them. See Fig.\ref{fig:extrapolation} for an illustrative example. 

\subsection{Continuation-Prompt Mixed Training}
Although framing zero-shot TTS as speech continuation effectively leverages the strong in-context learning capabilities of LLM and thus achieves high speaker similarity, it also introduces a degree of training/inference mismatch. During training, the emotion, semantics, prosody, and speaking style are usually consistent within an utterance, and therefore will be consistent between the reference speech and the generation target. However, during inference the semantics or emotion expressed by the reference utterance may be very different than those implicitly expressed in the target transcript input by a user. In this case, the model must be able to adjust the prosody and emotional delivery to match the target transcript, even if they do not match those found in the reference utterance. 
We propose a simple method to mitigate this mismatch: continuation-prompt mixed (CPM) training (left hand side of Fig.~\ref{fig:model}). During each training iteration, with some probability $p$ we randomly sample two different utterances from the same speaker and use one as the reference utterance and one as the target. With probability $1-p$ we use standard speech continuation based-training, i.e. we only sample one utterance and mask it, using the unmasked portion of the utterance as the reference and the masked portion of the utterance as the target. When using two different utterances we also have the opportunity to perform data augmentation to one but not the other. Specifically, we apply speed perturbation to the reference utterance with probability $p'$ and speed factor $\delta=0.25$. This randomly speeds up or slows down the reference speech, keeping its total duration within $1\pm \delta$ of its original duration. When a prompt is selected, we use learnable \texttt{sep\_token}s to separate the prompt and target, and only calculate the loss on target tokens.

\textbf{Prompt repetition for improving speaker similarity.} Another benefit of CPM training is that it allows us to improve speaker similarity between the generated speech and the reference speech by simply repeating the reference utterance and its transcript multiple times, without harming intelligibility. Counterintuitive as it might be, since no additional information is provided by repeating the prompt, we hypothesize that repeating the reference utterance increases the influence of the ground-truth speech tokens when performing the weighted sum over tokens within attention layers, biasing the model to attend more to the reference speech tokens than its previous generations, resulting in generated speech that is more similar to the reference prompt. 
As we quantitatively show in our experiments (Fig.~\ref{fig:repeatPrompt_F5}), when a model is trained solely on speech continuation, reference prompt repetition degrades the intelligibility (WER) of the generated speech because it biases the model to exhibit a repetitive style.
On the other hand, when using CPM training this issue is mitigated because 1) the model has been trained on examples where the reference speech and target speech express different styles and 2) \texttt{sep\_token}s explicitly separates the reference and target utterances, makes it easier to align the text and the corresponding speech.
\vspace{-0.2in}
\section{Experiments}
\vspace{-0.1in}
\subsection{Setup}
\textbf{Training data.}
Our training set consists of the English portion of Emilia~\cite{He2024EmiliaAE} and a subset of the training splits of Libriheavy and Libriheavy-long~\cite{Kang2023LibriheavyA5}, totaling 65K hours. Emilia is a recently proposed multilingual dataset comprised of in-the-wild speech of diverse styles. Its English portion contains 46K hours of speech, and after filtering out low quality data following~\cite{f5tts}, approximately 40K hours remain with a maximum utterance duration of 30 seconds. Libriheavy is an automatically transcribed version of the LibriLight audiobook corpus. The original release of Libriheavy contains 50k hours of speech, and we also make use of the recently released Libriheavy-long dataset, which is a re-segmented version that has longer segments ranging from 20 seconds to 100 seconds, totaling 42K hours. Note that in order to fairly compare with other models on extrapolation, we only train our models on utterances with duration less than or equal to 30 seconds. We randomly sampled 25K hours of speech from Libriheavy. Crucially, both Emilia and Libriheavy contain speaker information (although speaker information in Emilia is provided using automatic speaker diarization, we found it to be very reliable), and therefore we are able to sample multiple utterances from the same speaker for CPM training. We phonemize all text transcripts into the IPA phoneme set using espeak-ng~\cite{espeak-ng}.

\textbf{Evaluation tasks and data.}
We evaluate our models on two English-language, zero-shot TTS tasks: short-form TTS and long-form TTS. For short-form TTS ($\leq$20s): we use Seed-TTS eval set~\cite{Anastassiou2024SeedTTSAF} and Librispeech-PC~\cite{f5tts}, each containing around 1000 prompt-target pairs, where both the prompt and target speech are expected to be less than 10 seconds. For Long-form TTS, we consider three duration ranges (20-30s, 30-40s, and 40s-50s) and source the evaluation data from the libriheavy test and validation sets (we use validation set due to the scarcity of long-form speech samples in the test set. We thus avoid using the validation set to do any hyperparameter tuning or early stopping of model training). We sample 1000 prompt target pairs for the 20-30s range, 500 pairs for the 30-40s range, and 100 pairs for the 40-50s duration range. 
For our ablation studies, we sample a 1000 utterance evaluation set from the Libriheavy validation set with lengths shorter than 20 seconds for short-form ablations (Tab.~\ref{tab:ablation}), and lengths between 20 and 30 seconds for extrapolation (Tab.~\ref{tab:ablation-extrapolation}, no overlap with the testsets). For duration specification with the $\leq$20s test sets, we use follow~\cite{f5tts} and estimate the duration of the target sequence by calculating the speaking rate of the reference prompt (in terms of seconds-per-character) and multiplying this by the character length of the target text; for ablations and long-form TTS, we use the ground truth duration because the estimated duration sometimes falls outside of the target range (for example, when testing model performance on 40-50s speech, the estimated duration could be only 35 seconds), and we only compare against models that can also control the duration of their generations. We also show in Appendix~\ref{sec:dur-est} our model's performance on long-form TTS when using the automatically estimated durations.

\textbf{Model, training, inference, and baselines.}
Our main model has 840 million parameters, composed of 12 standard Transformer encoder layers and 40 Transformer decoder layers, where the dot-product attention mechanism is replaced by our proposed \pmrope. The hidden dimensions for both the encoder and decoder are 1024, and the model has 16 attention heads. For our ablation studies, we train 230 million parameter models with a hidden states dimension of 768, composed of 10 encoder layers and 16 decoder layers. The decoder-only model used in our ablation studies has 16 layers with a hidden dimension of 1024. We use the 4-codebook 50Hz Encodec model released by VoiceCraft~\cite{voicecraft} as our speech tokenizer. The pseudo sequence length $N$ is set to 2000 for all models where \pmrope is used. Models are trained with ScaledAdam~\cite{Yao2023ZipformerAF} with a maximum batch size of 0.3(ablation)/1.78(main) hours of audio. The base learning rate is $0.03$ and scheduled by the Eden scheduler~\cite{Yao2023ZipformerAF}. All models are trained for 50k steps with codebook loss weights of \{5, 1, 0.5, 0.1\}, similarly to~\cite{voicecraft}. Our main model is further trained for an additional 18k steps with codebook weights \{2.5, 2, 1.5, 0.6\}, as we found it to improve both intelligibility and speaker similarity metrics. Our main model is trained for 8 days on 8 L40 and 16 GH200 GPUs. The main model is trained on the entire 65k hour training set with maximal context length of 30 seconds, and the ablation models are trained on a 16k hour subset with maximal context length of 20 seconds. We use top-k sampling during inference with $k=10$.
For our baselines, we compare against VoiceCraft~\cite{voicecraft}, FireRedTTS~\cite{Guo2024FireRedTTSAF}, CosyVoice~\cite{Du2024CosyVoiceAS}, MaskGCT~\cite{f5tts}, F5-TTS~\cite{f5tts}, CosyVoice2~\cite{Du2024CosyVoice2S}, and Llasa~\cite{Ye2025LlasaST}. We additionally report in Appendix~\ref{sec:reported_reproduced} the performance of closed source models: Seed-TTS~\cite{Anastassiou2024SeedTTSAF}, and concurrent work Metis~\cite{Wang2025MetisAF} and SparkTTS~\cite{Wang2025SparkTTSAE}. 

\textbf{Evaluation Metrics.}
We use word error rate (WER) and speaker similarity (SpkSim) as automatic evaluation metrics following prior works. For our ablation studies, we additionally use UTMOS\cite{saeki2022utmosutokyosarulabvoicemoschallenge} to measure naturalness and DurDiff defined as the absolute difference between target generation duration and actual duration, to measure duration controllability. For comparison with other SoTA models on short-form TTS, we follow Seed-TTS eval setup and use Whisper-v3~\cite{Radford2022RobustSR} for ASR and the WavLM speaker verification model~\cite{Chen2021WavLMLS} for SpkSim. For long-form TTS, we used Whisper Large-v3-turbo for ASR as we found it to be both more accurate and faster. We additionally conduct subjective human evaluation using Amazon Mechanical Turk, focusing on three aspects: intelligibility, naturalness, and speaker similarity. For $\leq$20s testsets, since most models performance similarly, we adapt the Comparative MOS protocal from~\cite{loizou2011speech}, where we compare each generated sample with ground truth target sample on a 7-point Likert scale (-3 to 3). For 20-50s testsets, we use the regular MOS protocol, where we ask humans to rate each generated sample on a 5-point Likert scale (1 to 5). For each sample, we collect 10 human ratings for CMOS and 5 for MOS. To facilitate a fairer comparison, we also resample all speech waveforms to 16 kHz. 

\subsection{Ablations}\label{sec:ab}
\begin{table*}[ht]
\centering
\caption{Ablation studies on enc-dec architecutre, \pmrope, and CPM training. CPM stands for contiuation-prompt mixed training. SA stands for speech augmentation on prompt during training.}\label{tab:ablation}
% \resizebox{0.95\textwidth}{!}{
\begin{tabular}{lllrrrrrr}
\toprule
Architecture & PE & Training & WER $\downarrow$ & UTMOS $\uparrow$& SpkSim $\uparrow$ & DurDiff $\downarrow$\\
\midrule
Dec & Sinusoid & Continuation & 11.04 & 2.767 & 0.537  & 1.245 \\
Enc-Dec & RoPE & Continuation & 8.82 & 3.301 & 0.592 & 1.812 \\
Enc-Dec & \pmrope & Continuation & 8.49 & 3.337 & 0.601 & 0.009 \\
Enc-Dec & \pmrope & CPM & 6.42 & 3.365 & 0.587 & 0.009 \\
Enc-Dec & \pmrope & CPM+SA & 5.66 & 3.345 & 0.583 & 0.009 \\
\bottomrule
\end{tabular}
% }
\vspace{-0.2in}
\end{table*}

\begin{figure}[ht]
\vspace{-0.7in}
    \centering
    % Left: Figure (top-aligned)
    \begin{minipage}[t]{0.55\textwidth}
        \centering
        \includegraphics[width=\linewidth]{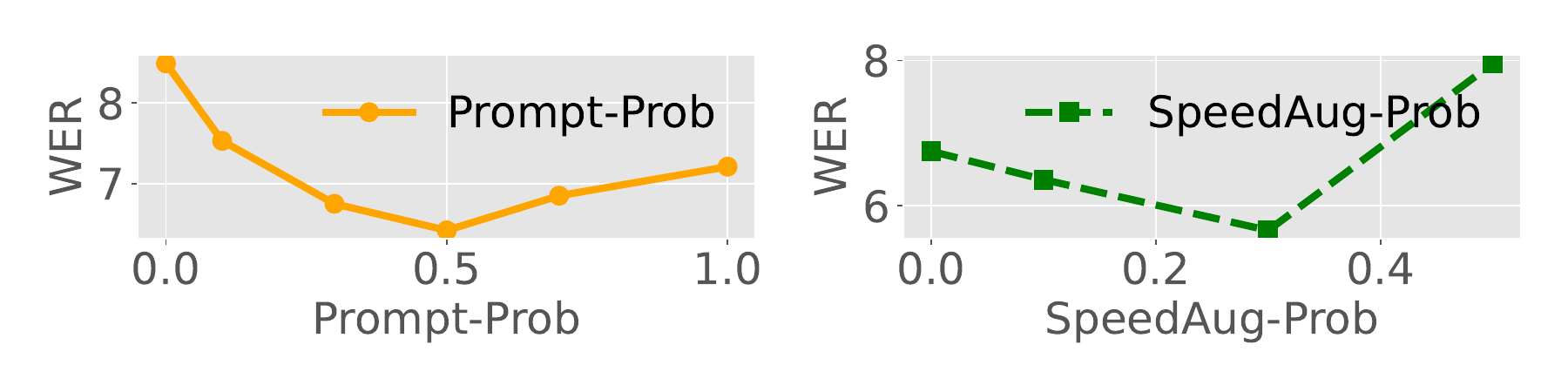}
        \captionof{figure}{Effects of prompt-prob and speed augmentation for CPM training.}
        \label{fig:ablation}
    \end{minipage}
    \hfill
    % Right: Table (also top-aligned)
    \begin{minipage}[t]{0.39\textwidth}
        \vspace*{-5em}  % Move the table up to align with the figure
        \centering
        \captionof{table}{\small Extrapolation capabilities of RoPE and \pmrope. max training context length: 20s, test: 20s to 30s.}
        \resizebox{1\textwidth}{!}{
        \label{tab:ablation-extrapolation}
        \begin{tabular}{llrr}
        \toprule
        Enc PE & Dec PE & WER$\downarrow$ & SpkSim $\uparrow$ \\
        \midrule
        RoPE & RoPE & 11.46 & 0.630  \\
        RoPE & \pmrope & 28.67 & 0.605  \\
        \pmrope & \pmrope & 6.75 & 0.640 \\
        \bottomrule
        \end{tabular}
        }
    \end{minipage}
    \vspace{-0.25in}
\end{figure}

Tab.~\ref{tab:ablation} shows how each component of our model contributes to its overall performance. We first compare the performance of the decoder-only model with sinusoidal position embedding (PE) and the encoder-decoder model with RoPE. The encoder-decoder model outperforms the decoder-only in all metrics except for DurDiff. But when RoPE is replaced by \pmrope (line 3 v.s. line 4), we see DurDiff drops from 1.812 to 0.009, which is a lower bound for this metric under our setup, because the codec token resolution is 0.02s per token. This illustrates the dramatic effectiveness of \pmrope~in enabling precise duration control. We also see a slight improvement on all other metrics, indicating the benefit of the improved text-speech alignment introduced by \pmrope. 
Lastly, we see further improvement when CPM and speed augmentation is applied. Fig.~\ref{fig:ablation} further shows the impact of different probability of using prompt as prefix and speed augmentation on prompt during training.
Tab.~\ref{tab:ablation-extrapolation} shows the benefit of \pmrope~on test time extrapolation, where the models trained on a maximal 20 seconds context length (prompt + target speech duration) are tested on 20 to 30 seconds context length. 
We see that to ensure extrapolation, it is necessary to apply \pmrope~on both the encoder and the decoder, additionally we see that \pmrope~improves the extrapolation WER from 11.46 to 6.75 compared to RoPE.

\begin{wrapfigure}{r}{0.4\textwidth}  % "r" for right, width can be adjusted
\vspace{-0.15in}
    \centering
    \includegraphics[width=1\linewidth]{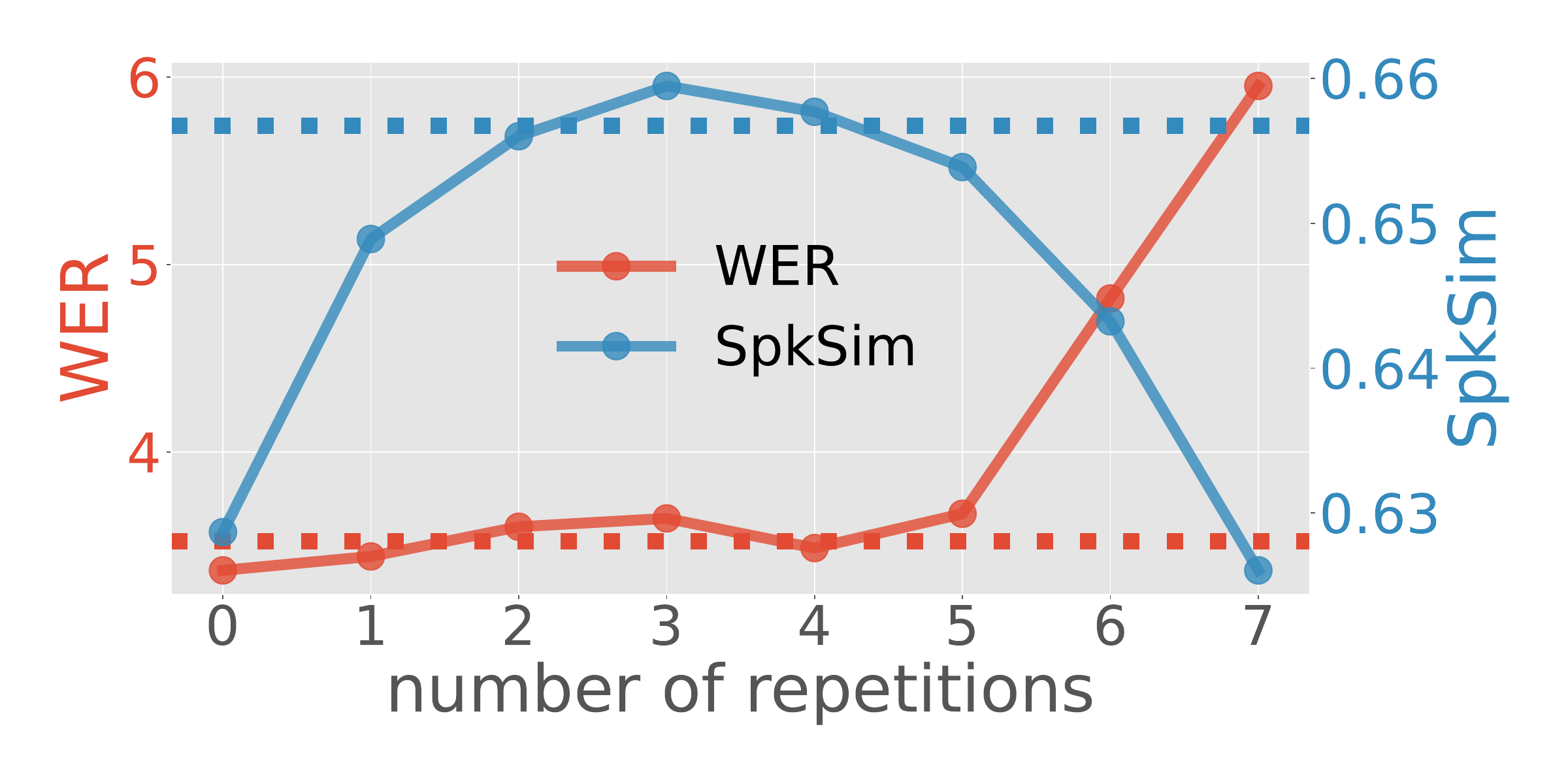}
    \caption{Ablation on the impact prompt repetition on generation quality measured by WER and SpkSim. The red and blue horizontal dotted lines are the performance when we repeat prompt so that the context length reach the maximal training duration.}
    \label{fig:repeatPrompt}
\end{wrapfigure}

Fig.~\ref{fig:repeatPrompt} shows the impact of prompt repetition on generation quality, measured by WER (left in red) and SpkSim (right in blue). 
The red and blue dotted horizontal lines indicate the performance when we repeat prompt for each sample so that the context length (i.e. prompt + generation length) reaches the maximum seen during training, i.e. 20 seconds.
We see that as the number of repetitions increases from 0 to 3, WER sees very little degradation, while SpkSim improves from around 0.63 to 0.66. But as we further increase the number of repetitions, both WER and SpkSim start to degrade. 
In our main experiments on short-form TTS, we choose to repeat each sample until the context length reaches the maximal training duration (i.e. the approach that produces the dotted lines in Figure\ref{fig:repeatPrompt}). On long-form TTS, we repeat the prompt once for the 20-30s range and do not apply prompt repetition to longer ranges. To test whether prompt repetition can also improve the performance of models trained with speech continuation, we tested F5-TTS on the same dataset using the same prompt repetition technique. 
We found that for F5-TTS, prompt repetition has a significantly negative impact on intelligibility: repeating prompt to reach the maximum duration seen during training duration degrades WER from 3.1 to 75.4. Further experiments on the impact of prompt repetition on F5-TTS can be found in Appendix~\ref{sec:repeatPrompt}. 

% \begin{figure}
%     \centering
%     \includegraphics[width=1\linewidth]{src/figs/repeatPrompt.pdf}
%     \caption{Ablation on the impact prompt repetition on generation quality measured by WER and SpkSim. The red and blue horizontal dotted lines are the performance when we repeat prompt so that the context length reach the maximal training duration.}
%     \label{fig:repeatPrompt}
% \end{figure}

\vspace{-0.1in}
\subsection{Short-form TTS}
\begin{table}[ht]
\vspace{-0.2in}
    \caption{Short context evaluation sets ($\leq20$ seconds): LibriSpeech-PC and Seed-TTS. N-CMOS stands for naturalness comparative MOS, and S-CMOS stands for Speaker Similarity Comparative MOS, where model generated samples are compared to ground truth. Except for ground truth which are taken from~\cite{Anastassiou2024SeedTTSAF} and~\cite{f5tts}, numbers for other models are reproduced by us using the corresponding official codebases, and a comparison between the reported numbers in other papers and our reproduced numbers can be found in App.~\ref{sec:reported_reproduced}.}\label{tab:short_tts}
    \centering
    % \resizebox{0.49\textwidth}{!}{
    \begin{tabular}{lrrrr}
    \toprule
    Model & WER & SpkSim & S-CMOS & N-CMOS  \\
    \midrule
    \multicolumn{5}{c}{\textbf{LibriSpeech-PC}} \\
    \midrule
    {\color{myLightGray} Ground Truth} & {\color{myLightGray} 2.23} & {\color{myLightGray} 0.69} & {\color{myLightGray} 0.00} & {\color{myLightGray} 0.00} \\
    FireRedTTS~\cite{Guo2024FireRedTTSAF} & 7.73 & 0.45 & -0.48{\tiny$\pm$0.22}& -0.89{\tiny$\pm$0.21}   \\
    Llasa 1B~\cite{Ye2025LlasaST} & 4.28 & 0.47 & 0.27{\tiny$\pm$0.19} & 0.13{\tiny$\pm$0.20}  \\
    CosyVoice~\cite{Du2024CosyVoiceAS} & 3.40 & 0.66 & 0.37{\tiny$\pm$0.19} & 0.07{\tiny$\pm$0.21}  \\
    VoiceCraft~\cite{voicecraft} & 3.09 & 0.52 & 0.15{\tiny$\pm$0.20} & -0.37{\tiny$\pm$0.20}  \\
    CosyVoice2~\cite{Du2024CosyVoice2S} & 2.66 & 0.66 & 0.40{\tiny$\pm$0.19} & 0.06{\tiny$\pm$0.21}  \\
    MaskGCT~\cite{Wang2024MaskGCTZT} & 2.66 & 0.66 & 0.21{\tiny$\pm$0.20} & -0.25{\tiny$\pm$0.21}  \\
    F5-TTS~\cite{f5tts} & 2.57 & 0.65 & 0.12{\tiny$\pm$0.20} & -0.43{\tiny$\pm$0.21}  \\
    \modelname & 2.64 & 0.63 & \textbf{0.60{\tiny$\pm$0.19}}& \textbf{0.18{\tiny$\pm$0.21}}  \\
    \midrule
    \multicolumn{5}{c}{\textbf{Seed-TTS (en)}} \\
    \midrule
    {\color{myLightGray} Ground Truth} & {\color{myLightGray} 2.15} & {\color{myLightGray} 0.73} & {\color{myLightGray} 0.00} & {\color{myLightGray} 0.00} \\
    FireRedTTS~\cite{Guo2024FireRedTTSAF} & 9.09 & 0.45 & -1.24{\tiny$\pm$0.16}& -0.92{\tiny$\pm$0.18}  \\
    Llasa 1B~\cite{Ye2025LlasaST} & 5.13 & 0.58 & -0.23{\tiny$\pm$0.14}& \textbf{0.05{\tiny$\pm$0.16}}  \\
    CosyVoice~\cite{Du2024CosyVoiceAS} & 4.20 & 0.64 & -0.42{\tiny$\pm$0.16}& -0.36{\tiny$\pm$0.18}  \\
    VoiceCraft~\cite{voicecraft} & 3.45 & 0.52 & -0.84{\tiny$\pm$0.19}& -0.73{\tiny$\pm$0.18}  \\
    CosyVoice2~\cite{Du2024CosyVoice2S} & 2.69 & 0.66 & \textbf{-0.11{\tiny$\pm$0.16}}&  -0.19{\tiny$\pm$0.15}  \\
    MaskGCT~\cite{Wang2024MaskGCTZT} & 2.52 & 0.71 & -0.13{\tiny$\pm$0.15}& -0.14{\tiny$\pm$0.15}  \\
    F5-TTS~\cite{f5tts} & 1.78 & 0.66 & -0.18{\tiny$\pm$0.14}& -0.21{\tiny$\pm$0.16}  \\
    \modelname & 2.15 & 0.63 & -0.32{\tiny$\pm$0.15}&  -0.18{\tiny$\pm$0.14}  \\
    \bottomrule
    \end{tabular}
    % }
    \vspace{-0.1in}
\end{table}
As shown in Tab.~\ref{tab:short_tts}, on Librispeech-PC~\cite{f5tts}, \modelname~outperforms other models during human evaluation on both naturalness (N-CMOS) and speaker similarity (S-CMOS). Notably, \modelname~(along with several other models) even outperforms the ground truth speech. Anecdotally, we hypothesize that this is due to the fact that audiobook speech is highly regular and sometimes monotonic, but since our model is also trained on in-the-wild conversational data, it tends to generate speech with a more natural flow. 
On Seed-TTS (en)~\cite{Anastassiou2024SeedTTSAF}, \modelname~is slightly worse than the best performing models Llasa 1B and MaskGCT. However, we find that the automatic WER and SpkSim metrics are not perfectly correlated with human judgements: for example, Llasa 1B achieves a SpkSim of 0.58, which is lower than \modelname, but humans rate it to have a higher speaker similarity than \modelname. Overall, most models perform similarly on the two test sets and are often rated as equal or even better than ground truth.

\subsection{Long-form TTS}
\begin{table}[H]
\vspace{-0.2in}
    \caption{Long context evaluation results. All models are trained with a maximal training length of 30 seconds. I in I-MOS stands for intelligibility, N stands for naturalness, and S stands for speaker similarity.}\label{tab:long_tts}
    \centering
    % \resizebox{0.49\textwidth}{!}{
    \begin{tabular}{lrrrrr}
    \toprule
    Model & WER & SpkSim & S-MOS & I-MOS & N-MOS  \\
    \midrule
    \multicolumn{6}{c}{\textbf{20s-30s, within training duration}} \\
    \midrule
    {\color{myLightGray} Ground Truth} 
    & {\color{myLightGray} 3.78} 
    & {\color{myLightGray} 0.86}
    & {\color{myLightGray} 3.95{\tiny$\pm$0.16}}& {\color{myLightGray} 4.01{\tiny$\pm$0.13}}
    & {\color{myLightGray} 3.63{\tiny$\pm$0.14}}
     \\
    F5-TTS~\cite{f5tts} & 5.31 & 0.70 & 4.03{\tiny$\pm$0.15}& 3.47{\tiny$\pm$0.15} & 2.97{\tiny$\pm$0.17}  \\
    MaskGCT~\cite{Wang2024MaskGCTZT} & 3.91 & 0.76 & \textbf{4.29{\tiny$\pm$0.13}}& 3.55{\tiny$\pm$0.16} & 3.07{\tiny$\pm$0.18}  \\
    \modelname & 3.53 & 0.72 & 4.13{\tiny$\pm$0.14}& \textbf{3.75{\tiny$\pm$0.14}} & \textbf{3.45{\tiny$\pm$0.15}}  \\
    \midrule
    \multicolumn{6}{c}{\textbf{30s-40s, extrapolation}} \\
    \midrule
    {\color{myLightGray} Ground Truth} & {\color{myLightGray} 3.13} & {\color{myLightGray} 0.86}& {\color{myLightGray} 4.12{\tiny$\pm$0.19}} & {\color{myLightGray} 4.21{\tiny$\pm$0.15}} & {\color{myLightGray} 4.05{\tiny$\pm$0.16}}  \\
    F5-TTS~\cite{f5tts} & 34.15 & 0.70 & 3.97{\tiny$\pm$0.15}& 2.21{\tiny$\pm$0.15} & 2.15{\tiny$\pm$0.16}  \\
    MaskGCT~\cite{Wang2024MaskGCTZT} & 13.81 & 0.75 & \textbf{4.27{\tiny$\pm$0.14}}& 3.06{\tiny$\pm$0.18} & 2.79{\tiny$\pm$0.18}  \\
    \modelname & 7.27 & 0.70 & 4.11{\tiny$\pm$0.17}& \textbf{4.16{\tiny$\pm$0.14}} & \textbf{3.69{\tiny$\pm$0.16}}  \\
    \midrule
    \multicolumn{6}{c}{\textbf{40s-50s, extrapolation}} \\
    \midrule
    {\color{myLightGray} Ground Truth} & {\color{myLightGray} 2.52} & {\color{myLightGray} 0.87}& {\color{myLightGray} 4.33{\tiny$\pm$0.16}} & {\color{myLightGray} 4.52{\tiny$\pm$0.11}} & {\color{myLightGray} 3.99{\tiny$\pm$0.16}}  \\
    F5-TTS~\cite{f5tts} & 52.44 & 0.70 & \textbf{4.04{\tiny$\pm$0.17}}& 1.94{\tiny$\pm$0.15} & 2.35{\tiny$\pm$0.14} \\
    MaskGCT~\cite{Wang2024MaskGCTZT} & 82.29 & 0.65 & 3.95{\tiny$\pm$0.20}& 1.42{\tiny$\pm$0.10} & 1.61{\tiny$\pm$0.13}  \\
    \modelname & 11.91 & 0.70 & 3.94{\tiny$\pm$0.18}& \textbf{3.23{\tiny$\pm$0.19}} & \textbf{3.23{\tiny$\pm$0.19}}  \\
    \bottomrule
    \end{tabular}
    % }
\end{table}
Long-form TTS is much more challenging than short-form. To force a model to generate speech of a predefined duration, we only compare models that support duration control. We see in Tab.~\ref{tab:long_tts} that across different context length ranges, \modelname~significantly outperforms other models on WER, intelligibility MOS (I-MOS), and naturalness MOS (N-MOS), and the gap is especially large as the context length increases. As for speaker similarity, \modelname~is a close second on every range, and is on par with or better than ground truth target speech in the 20-40s range. By listening to bad samples, we found that different models have different failure modes - for F5-TTS and MaskGCT, the models tend to clone the prompt voice well but devolve into unintelligible speech or random words. In contrast, \modelname~sometimes generates speech in a voice that that slightly deviates from that of the prompt, but the generated words follow the target transcript closely and with a high naturalness.

\section{Conclusion}
We introduce \modelname, a novel autoregressive NCLM-based, zero-shot TTS model that achieves robust and duration-controllable speech synthesis, as well as the ability to generate speech longer than what was seen during training. The key novel contributions of our approach include the Progress-Monitoring Rotary Position Embedding (\pmrope) for effective text-speech alignment, duration control, and extrapolation; and continuation-prompt mixed training (CPM) to address the training/inference mismatch. By scaling up the training data and model size, \modelname~sets new state-of-the-art results on both short-form and long-form TTS benchmarks.

% \section{Acknowledgements}
 % excluded in the anonymous version

% \bibliographystyle{acl_natbib}
% \newpage
\bibliographystyle{unsrt}
\bibliography{voicestar}

\appendix

\newpage
\appendix
\section{Model Details}\label{sec:app-detail}
\subsection{RoPE and PM-RoPE for higher dimensions}
For higher dimensions, rotation matrix is defined by dividing the space into D/2 2-dimensional spaces, and applying a 2-dimensional rotation matrix (with different $\theta$s) to each space. Mathematically, see Eq.~\ref{fn:rope-RMat}. Similarly, the rotation matrix for PM-RoPE in higher dimension is show in Eq.~\ref{fn:pmrope-RMat}, where $T$ is the total length of the sequence.
% \onecolumn
\begin{equation}
    \resizebox{0.8\textwidth}{!}{$   
	R^D_{\Theta,t} = 
	\begin{pmatrix}
		\cos{t\theta_1}& -\sin{t\theta_1}&0&0&\cdots&0&0\\
		\sin{t\theta_1}&\cos{t\theta_1}&0&0&\cdots&0&0 \\
		0&0&\cos{t\theta_2}& -\sin{t\theta_2}&\cdots&0&0\\
		0&0&\sin{t\theta_2}&\cos{t\theta_2}&\cdots&0&0 \\
		\vdots&\vdots&\vdots&\vdots&\ddots&\vdots&\vdots\\
		0&0&0&0&\cdots&\cos{t\theta_{D/2}}& -\sin{t\theta_{D/2}}\\
		0&0&0&0&\cdots&\sin{t\theta_{D/2}}&\cos{t\theta_{D/2}}
	\end{pmatrix}
	\label{fn:rope-RMat}
    $}
\end{equation}

\begin{equation}
    \resizebox{0.8\textwidth}{!}{$   
	R^D_{\Theta,t, T} = 
	\begin{pmatrix}
		\cos{\frac{t}{T}N\theta_1}& -\sin{\frac{t}{T}N\theta_1}&0&0&\cdots&0&0\\
		\sin{\frac{t}{T}N\theta_1}&\cos{\frac{t}{T}N\theta_1}&0&0&\cdots&0&0 \\
		0&0&\cos{\frac{t}{T}N\theta_2}& -\sin{\frac{t}{T}N\theta_2}&\cdots&0&0\\
		0&0&\sin{\frac{t}{T}N\theta_2}&\cos{\frac{t}{T}N\theta_2}&\cdots&0&0 \\
		\vdots&\vdots&\vdots&\vdots&\ddots&\vdots&\vdots\\
		0&0&0&0&\cdots&\cos{\frac{t}{T}N\theta_{D/2}}& -\sin{\frac{t}{T}N\theta_{D/2}}\\
		0&0&0&0&\cdots&\sin{\frac{t}{T}N\theta_{D/2}}&\cos{\frac{t}{T}N\theta_{D/2}}
	\end{pmatrix}
	\label{fn:pmrope-RMat}
    $}
\end{equation}
% \twocolumn
where $\Theta = \{\theta_i=10000^{-2(i-1)/D}, i \in [1, 2, ..., D/2]\}$. RoPE requires D is even.

\subsection{text-speech alignment in different models}\label{sec:alignment}
for decoder only model, text tokens usually preceed speech tokens; for encoder-decoder model, text tokens are the input to the encoder and speech tokens are the input to decoder, and speech token attends to text tokens via cross-attention. Here we visualize self-attention in encoder only model and cross-attention in encoder-decoder model. We set text and speech token embeddings to unit vectors, so that visualized attention maps solely reveal the inductive bias introduce by model architecture and position embeddings.

\begin{figure*}
    \centering
    \includegraphics[width=0.8\linewidth]{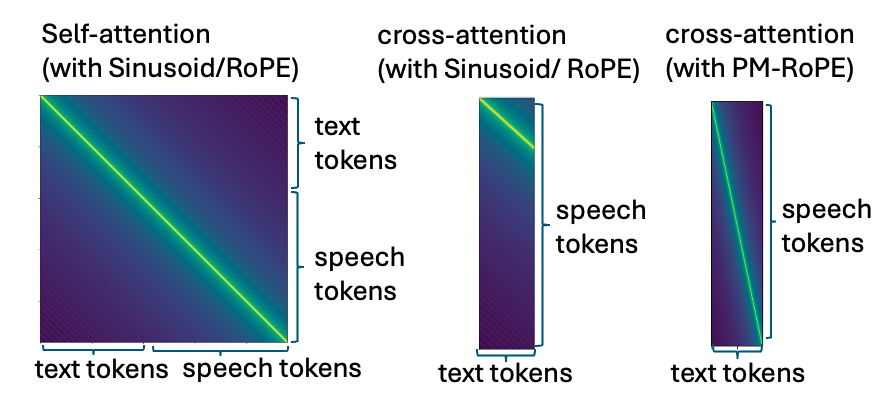}
    \caption{attention maps}
    \label{fig:alignment_attention}
\end{figure*}

\section{Limitations}
\textbf{Speaker similarity}. 
On short-form TTS benchmarks, thanks to the prompt repetition trick, our model is either SotA or on par with SotA on speaker similarity. However, on long-form TTS, since the context length is already at or exceeds maximal training duration, excessive prompt repetition can significantly degrade intelligibility and therefore our model exhibits a gap with SotA speaker similarity. We argue that speaker similarity is not a weakness of the main ideas proposed in this paper, and can be separately addressed by e.g. using more advanced neural codec model~\cite{Ye2025LlasaST}, or adapting multistage modeling approaches~\cite{Wang2024MaskGCTZT,Du2024CosyVoice2S}.

\textbf{Generation speed}. 
Our 840M model has a real-time factor (RTF) larger than 1 and therefore cannot perform faster-than-real-time generation. Although this could be mitigated by techniques such as quantization~\cite{dettmers2022llmint88bitmatrixmultiplication}, grouped prediction~\cite{chen2024valle2neuralcodec}, speculative decoding~\cite{li2025fast,nguyen2025accelerating}, etc.
\section{More Results}
\subsection{Comparing impact of prompt repetition on \modelname~v.s. F5-TTS}\label{sec:repeatPrompt}
\begin{figure*}[ht]
    \centering
    % First subfigure
    \begin{subfigure}[b]{0.48\linewidth}
        \centering
        \includegraphics[width=\linewidth]{src/figs/repeatPrompt.pdf}
        \caption{\modelname (same figure shown in main text).}
        \label{fig:repeatPrompt_app}
    \end{subfigure}
    \hfill
    % Second subfigure
    \begin{subfigure}[b]{0.48\linewidth}
        \centering
        \includegraphics[width=\linewidth]{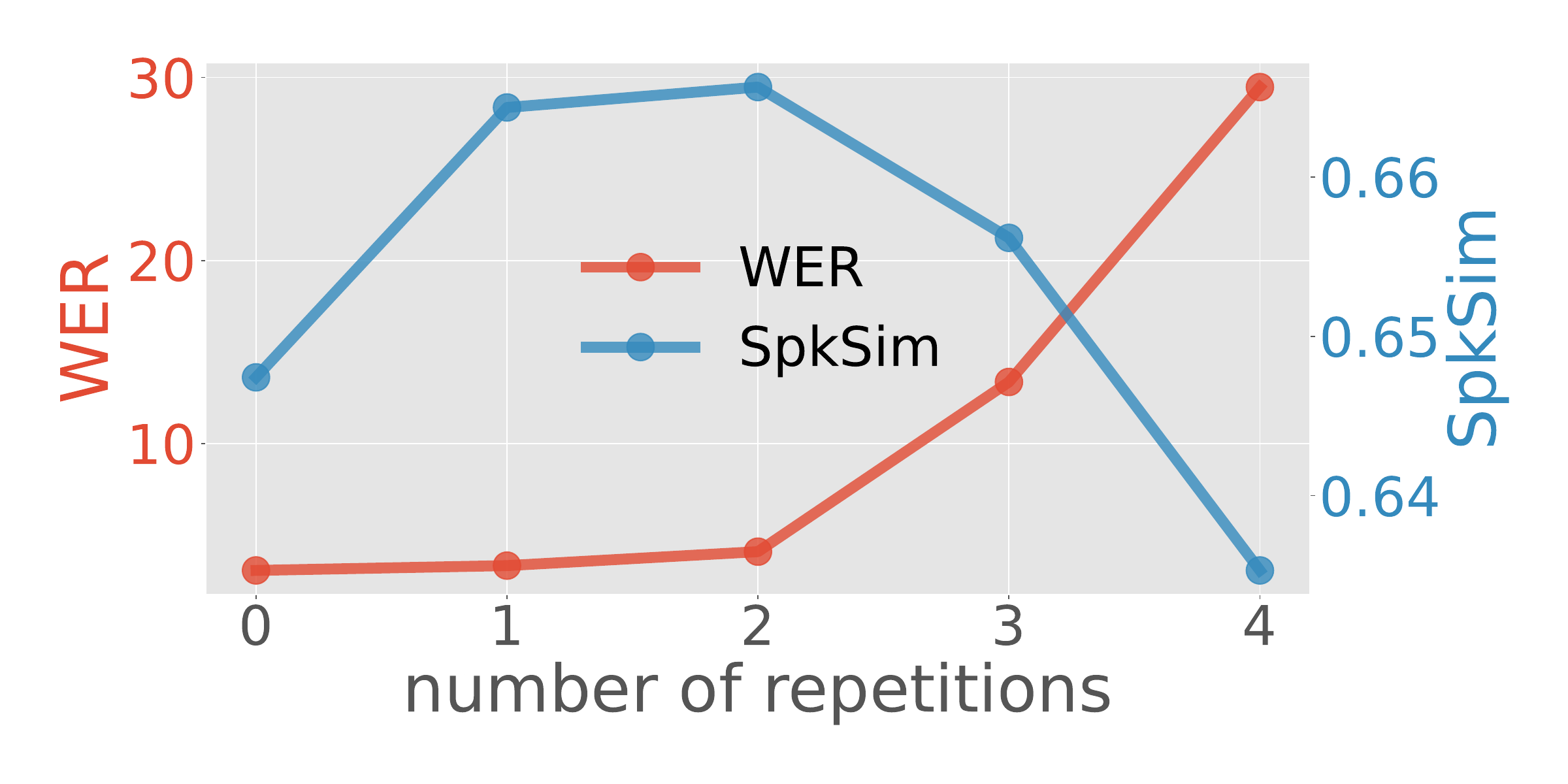}
        \caption{F5-TTS.}
        \label{fig:repeatPrompt_F5}
    \end{subfigure}

    \caption{Comparison of the impact of prompt repetition on \modelname~(left) and F5-TTS (right). In the \modelname~figure, the red and blue horizontal dotted lines indicate performance when we repeat the prompt so the context length (prompt + generation) reaches the maximal training duration. We do not show the dotted lines for F5-TTS because the model does not produce intelligible speech (WER is 75.4). We see that \modelname's WER stays stably low while SpkSim benefits from repeating the prompt until repeating for more than 6 times, while for F5-TTS, repeating the prompt for more than 2 times degrades WER significantly. Note that the WER y-axis ticks for \modelname~and F5-TTS are at very different scale.}
    \label{fig:repeatPrompt_combined}
\end{figure*}

Fig.~\ref{fig:repeatPrompt_app} and Fig.~\ref{fig:repeatPrompt_F5} shows the effect of prompt repetition technique on \modelname~and F5-TTS. We see that prompt repetition is not suitable for F5-TTS, because although it improves speaker similarity slightly when the number of repetitions is small, it quickly degrade WER significantly as the number of repetition increases.

\subsection{reported and reproduced results}\label{sec:reported_reproduced}
Tab.~\ref{tab:ls_reprod} and Tab.~\ref{tab:seed_reprod} show the reported and reproduced results of different models on Librisspeech-PC and Seed-TTS (en). We see that most reported numbers are closely reproduced, except for FireRedTTS and VoiceCraft. For VoiceCraft, we found that instead of using the recommended top p sampling with 0.9 probability, using top k sampling with top k of 40 significantly improves results.
\begin{table}[ht]
    \caption{LibriSpeech-PC~\cite{f5tts}. Reported numbers are taken from~\cite{f5tts}.}\label{tab:ls_reprod}
    \centering
    % \resizebox{0.5\textwidth}{!}{
    \begin{tabular}{lcc}
    \toprule
    Model &  WER & SpkSim \\
    \midrule
    \multicolumn{3}{c}{\textbf{Reported}} \\
    \midrule
    Ground Truth & 2.23 & 0.69 \\
    FireRedTTS &3.82 & 0.46 \\
    CosyVoice & 3.39 & 0.64 \\
    F5-TTS  & 2.42 & 0.66 \\
    \midrule
    \multicolumn{3}{c}{\textbf{Reproduced}} \\
    \midrule
    FireRedTTS & 7.73 & 0.45 \\
    Llasa 1B & 4.28 & 0.47 \\
    CosyVoice & 3.40 & 0.66 \\
    VoiceCraft & 3.09 & 0.52 \\
    MaskGCT & 2.66 & 0.66 \\
    CosyVoice2 & 2.66 & 0.66 \\
    F5-TTS  & 2.57 & 0.65 \\
    \modelname & 2.64 & 0.63 \\
    \bottomrule
    \end{tabular}
    % }
\end{table}
\begin{table}[ht]
    \caption{Seed-TTS English. Reported numbers are taken from the corresponding cited papers: Ground truth~\cite{Anastassiou2024SeedTTSAF}, Seed-TTS~\cite{Anastassiou2024SeedTTSAF}, VoiceCraft~\cite{Wang2024MaskGCTZT}, FireRedTTS~\cite{f5tts}, CosyVoice~\cite{Wang2024MaskGCTZT}, MaskGCT~\cite{Wang2024MaskGCTZT}, F5-TTS~\cite{f5tts}, CosyVoice2~\cite{Du2024CosyVoice2S}, Llasa 1B~\cite{Ye2025LlasaST}, Metis~\cite{Wang2025MetisAF}, SparkTTS~\cite{Wang2025SparkTTSAE}.
    For VoiceCraft, our reproduced version uses topk of 40 for sampling which leads to better performance. Models are ordered by arxiv submission time.}\label{tab:seed_reprod}
    \centering
    % \resizebox{0.5\textwidth}{!}{
    \begin{tabular}{lcc}
    \toprule
    Model &  WER & SpkSim\\
    \midrule
    \multicolumn{3}{c}{\textbf{Reported}} \\
    \midrule
    Ground Truth & 2.15 & 0.73 \\
    VoiceCraft & 7.56 & 0.47 \\
    FireRedTTS &3.82 & 0.46 \\
    CosyVoice & 3.39 & 0.64 \\
    Llasa 1B & 3.22 & 0.57 \\
    MaskGCT &2.47 & 0.72 \\
    CosyVoice2  & 2.38 & 0.65 \\
    Metis & 2.28 & 0.72 \\
    Seed-TTS  & 2.14 & 0.76 \\
    SparkTTS & 1.98 & 0.58 \\
    F5-TTS  & 1.83 & 0.67 \\
    \midrule
    \multicolumn{3}{c}{\textbf{Reproduced}} \\
    \midrule
    FireRedTTS & 9.09 & 0.45 \\
    Llasa 1B & 5.13 & 0.58 \\
    CosyVoice & 4.20 & 0.64 \\
    VoiceCraft & 3.45 & 0.52 \\
    CosyVoice2 & 2.69 & 0.66 \\
    MaskGCT & 2.52 & 0.71 \\
    F5-TTS  & 1.78 & 0.66 \\
    \modelname & 2.15 & 0.63 \\
    \bottomrule
    \end{tabular}
    % }
\end{table}

\subsection{Using Character duration-based estimation v.s. ground truth duration}\label{sec:dur-est}
Tab.~\ref{tab:gt-est} shows objective metrics comparing model performance using ground truth duration versus using estimated duration~\cite{f5tts}. We see that in most cases, using estimated duration will degrade the performance. We argue that ground truth duration is not necessary for good performance in that we could train duration predictor~\cite{Le2023VoiceboxTM,Lee2024DiTToTTSDT} that can potentially achieve much better duration estimation than the current naive approach.
\begin{table}[ht]
    \caption{Long context evaluation results listed as given duration estimated by character duration of prompt versus ground truth duration. All models are trained with a maximal training length of 30 seconds. I in I-MOS stands for intelligibility, N stands for naturalness, and S stands for speaker similarity.}\label{tab:gt-est}
    \centering
    % \resizebox{0.49\textwidth}{!}{
    \begin{tabular}{lcccc}
    \toprule
    Model & \multicolumn{2}{c}{WER} & \multicolumn{2}{c}{SpkSim} \\
    \midrule
    Duration    & GT & Est. &GT & Est. \\
    \midrule
    \multicolumn{5}{c}{\textbf{20s-30s, within training duration}} \\
    \midrule
    {\color{myLightGray} Ground Truth} 
    & {\color{myLightGray} 3.78}  & {\color{myLightGray} -} 
    & {\color{myLightGray} 0.86} & {\color{myLightGray} -}
     \\
    F5-TTS~\cite{f5tts} & 5.31 & 7.94& 0.70 & 0.70  \\
    MaskGCT~\cite{Wang2024MaskGCTZT} & 3.91& 4.74 & 0.76& 0.76   \\
    \modelname & 3.53& 4.60 & 0.72& 0.71   \\
    \midrule
    \multicolumn{5}{c}{\textbf{30s-40s, extrapolation}} \\
    \midrule
    {\color{myLightGray} Ground Truth} & {\color{myLightGray} 3.13}& {\color{myLightGray} -} & {\color{myLightGray} 0.86}& {\color{myLightGray} -}  \\
    F5-TTS~\cite{f5tts} & 34.15& 33.34 & 0.70& 0.70 \\
    MaskGCT~\cite{Wang2024MaskGCTZT} & 13.81& 26.39 & 0.75& 0.73   \\
    \modelname & 7.27& 8.29 & 0.70& 0.69   \\
    \midrule
    \multicolumn{5}{c}{\textbf{40s-50s, extrapolation}} \\
    \midrule
    {\color{myLightGray} Ground Truth} & {\color{myLightGray} 2.52}& {\color{myLightGray} -} & {\color{myLightGray} 0.87}& {\color{myLightGray} -}  \\
    F5-TTS~\cite{f5tts} & 52.44& 50.28 & 0.70& 0.70  \\
    MaskGCT~\cite{Wang2024MaskGCTZT} & 82.29& 77.24 & 0.65& 0.64   \\
    \modelname & 11.91& 17.33 & 0.70& 0.66  \\
    \bottomrule
    \end{tabular}
    % }
\end{table}
% \subsection{Short-form TTS}
\section{Instructions for human listening test}\label{sec:amt-instruction}
\begin{figure*}
    \centering
    \includegraphics[width=1\linewidth]{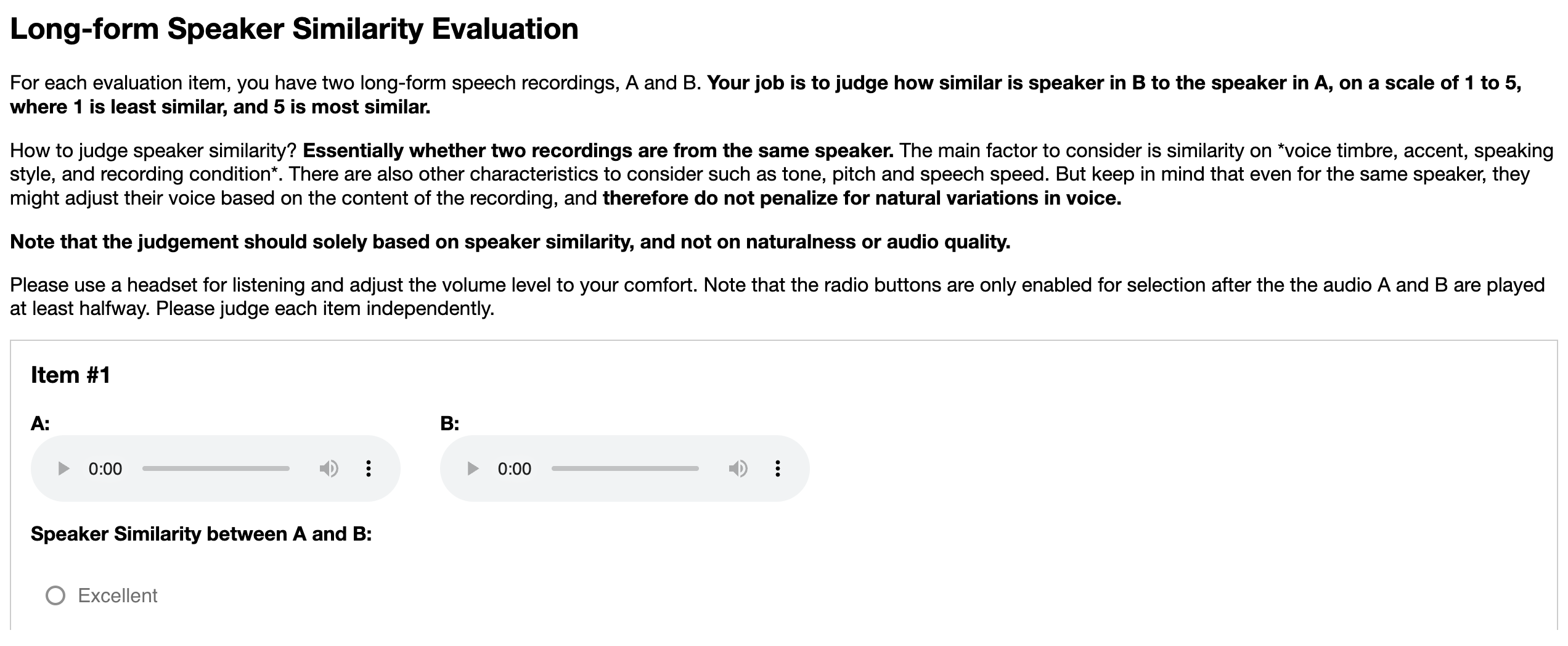}
    \caption{screenshots for speech intelligibility human evaluation. There are five options, Excellent, Good, Fair, Poor, Bad}\label{fig:amt_spk_sim}
\end{figure*}

Screenshots of instructions for all  human listening test we used on Amazon Mechanical Turk is shown in figure~\ref{fig:amt_comparative_nat} to figure~\ref{fig:amt_spk_sim} 
We used 13 dollar per hour to reward the workers. We only allow Turkers who are resident of the US, Australia, Canada, and UK to participate, in hope that the workers are native English speakers. We acknowledge that this is not a perfect approach and might need to bias in judgment, but since Amazon Mechanical Turk doesn't allow selection on native language, this is the best approach we could think of as a proxy to constraining the native language.
\begin{figure*}
    \centering
    \includegraphics[width=1\linewidth]{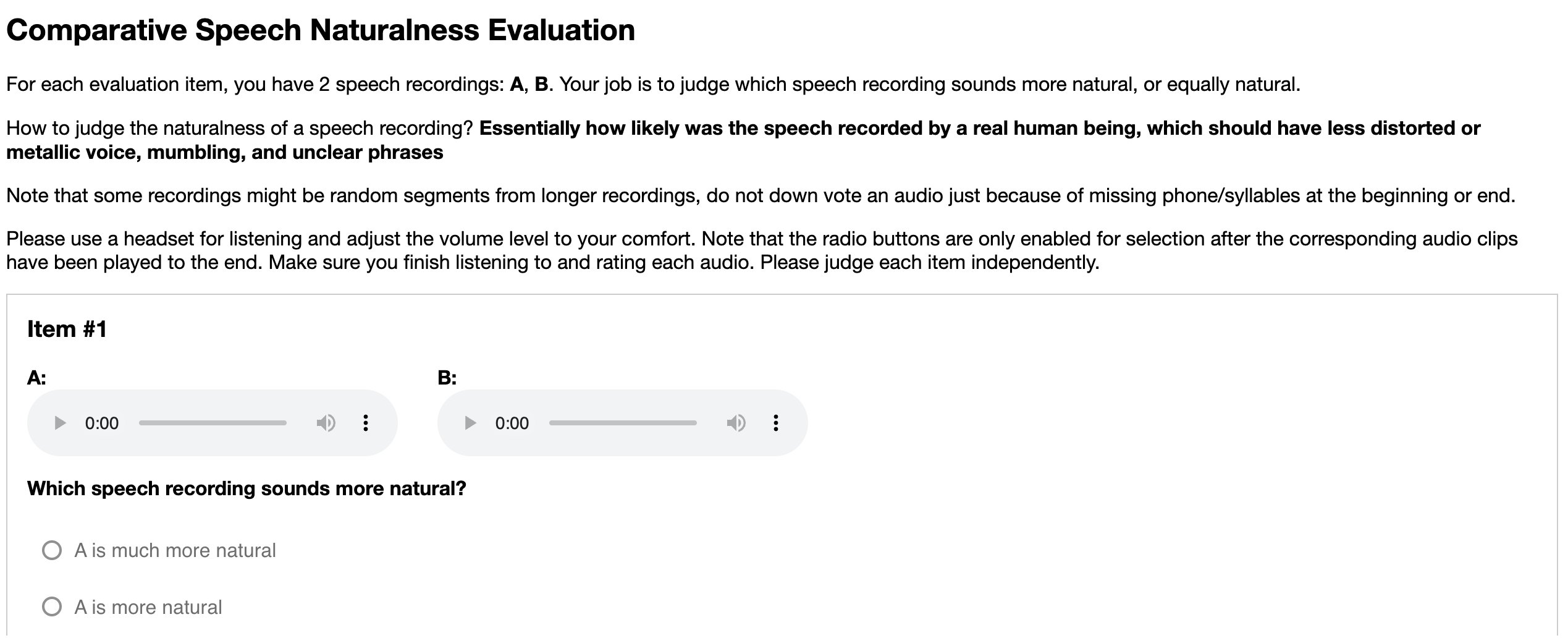}
    \caption{screenshots for comparative speech naturalness human evaluation. There are seven options: A is much more natural, A is more natural, A is slightly more natural, equally natural, B is much more natural, B is more natural, B is slightly more natural.}\label{fig:amt_comparative_nat}
\end{figure*}

\begin{figure*}
    \centering
    \includegraphics[width=1\linewidth]{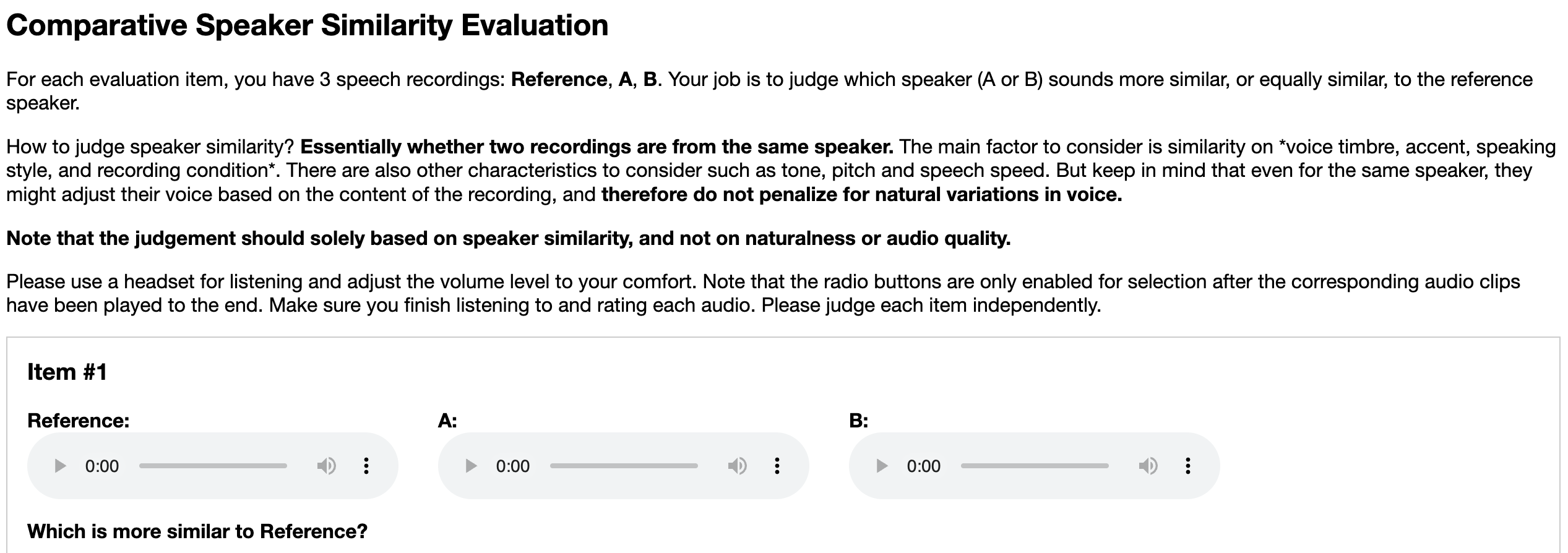}
    \caption{screenshots for comparative speaker similarity human evaluation. There are also seven options, replacing the word ``natural'' in comparative naturalness task with ``similar''}\label{fig:amt_comparative_nat}
\end{figure*}

\begin{figure*}
    \centering
    \includegraphics[width=1\linewidth]{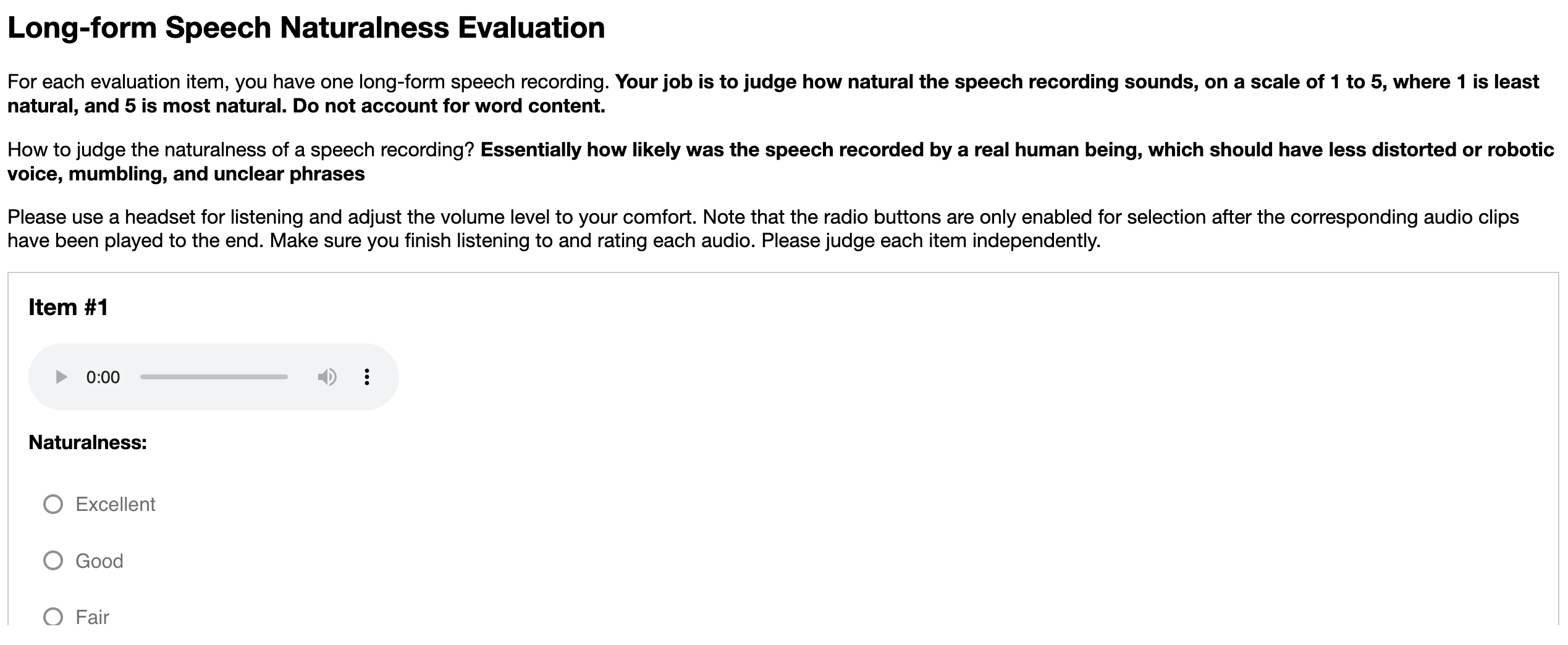}
    \caption{screenshots for speech naturalness human evaluation. There are five options, Excellent, Good, Fair, Poor, Bad}\label{fig:amt_nat}
\end{figure*}

\begin{figure*}
    \centering
    \includegraphics[width=1\linewidth]{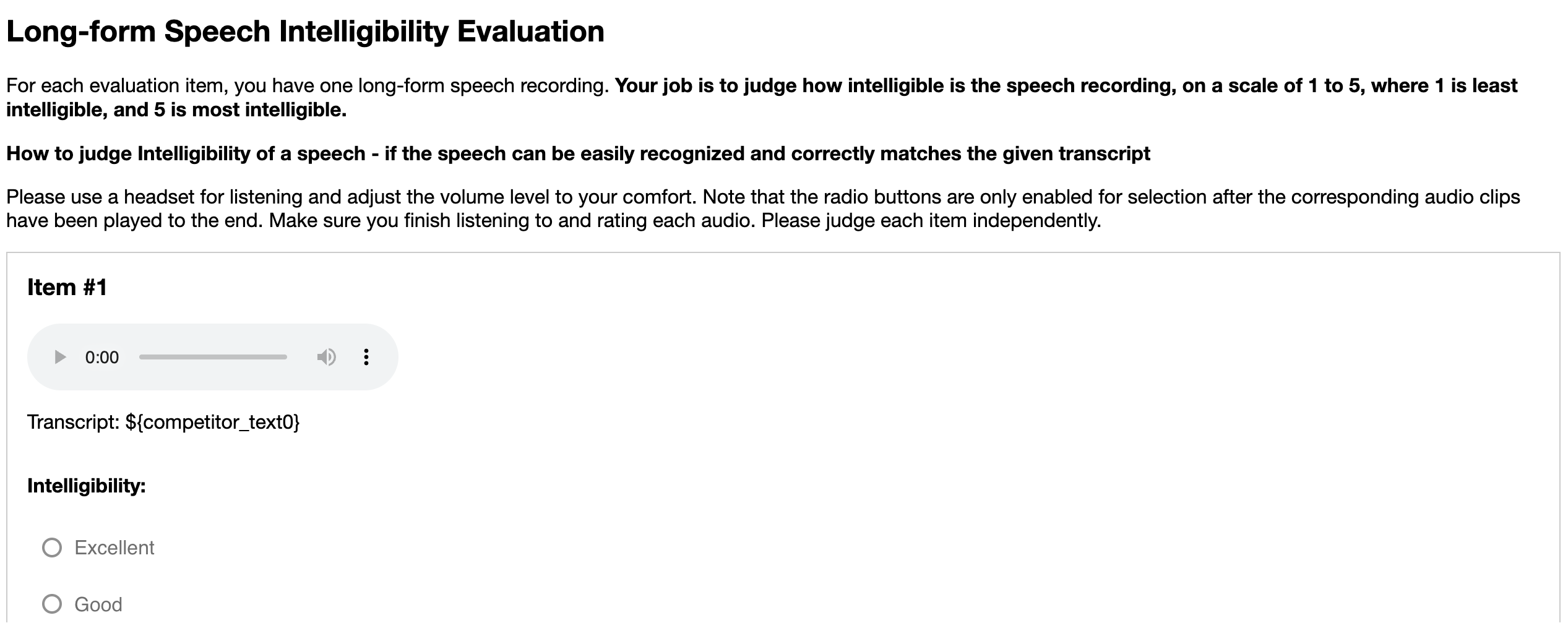}
    \caption{screenshots for speech intelligibility human evaluation. There are five options, Excellent, Good, Fair, Poor, Bad}\label{fig:amt_intelli}
\end{figure*}

\newpage

\end{document}